\documentclass[runningheads]{llncs}
\usepackage[T1]{fontenc}
\usepackage{graphicx}
\usepackage{booktabs} 
\usepackage{array}
\newcolumntype{P}[1]{>{\raggedright\arraybackslash}p{#1}}

\usepackage{caption}
\usepackage{subcaption}
\usepackage{orcidlink}
\usepackage{adjustbox}
\usepackage{multirow}
\hypersetup{
    colorlinks=true,
    linkcolor=blue,
    urlcolor=blue,
    citecolor=black,
    filecolor=black,
    linkbordercolor=white, 
    urlbordercolor=white,
    citebordercolor=white,
    filebordercolor=white,
}

\begin{document}
\title{Promptable Longitudinal Lesion Segmentation in Whole-Body CT}

%
%
\author{Yannick Kirchhoff\inst{1,2,3}\orcidlink{0000-0001-8124-8435} \and
        Maximilian Rokuss\inst{1,2,3}\orcidlink{0009-0004-4560-0760} \and 
        Fabian Isensee\inst{1,4}\orcidlink{0000-0002-3519-5886} \and
        Klaus H. Maier-Hein\inst{1,4,5}\orcidlink{0000-0002-6626-2463}
}
\authorrunning{Y. Kirchhoff et al.}

\institute{
German Cancer Research Center (DKFZ) Heidelberg,\\Division of Medical Image Computing, Heidelberg, Germany \and
HIDSS4Health - Helmholtz Information and Data Science School for Health, Karlsruhe/Heidelberg, Germany \and
Faculty of Mathematics and Computer Science,\\Heidelberg University, Heidelberg, Germany \and
Helmholtz Imaging, DKFZ, Heidelberg, Germany \and
Pattern Analysis and Learning Group, Department of Radiation Oncology, Heidelberg University Hospital, Heidelberg, Germany
\email{yannick.kirchhoff@dkfz-heidelberg.de}
}
\maketitle              
\begin{abstract}
Accurate segmentation of lesions in longitudinal whole-body CT is essential for monitoring disease progression and treatment response. While automated methods benefit from incorporating longitudinal information, they remain limited in their ability to consistently track individual lesions across time. Task 2 of the autoPET/CT IV Challenge addresses this by providing lesion localizations and baseline delineations, framing the problem as longitudinal promptable segmentation. In this work, we extend the recently proposed LongiSeg framework with promptable capabilities, enabling lesion-specific tracking through point and mask interactions. To address the limited size of the provided training set, we leverage large-scale pretraining on a synthetic longitudinal CT dataset. Our experiments show that pretraining substantially improves the ability to exploit longitudinal context, yielding an improvement of up to 6 Dice points compared to models trained from scratch. These findings demonstrate the effectiveness of combining longitudinal context with interactive prompting for robust lesion tracking. Code is publicly available at \url{https://github.com/MIC-DKFZ/LongiSeg/tree/autoPET}.
\end{abstract}

\section{Introduction}

Accurate lesion segmentation in whole-body CT is crucial for reliable monitoring of tumor progression and treatment response in oncology. While deep learning-based methods show strong performance in cross-sectional lesion segmentation~\cite{isensee2021nnu}, tracking individual lesions across timepoints remains a challenging task due to anatomical changes, differences in patient positioning, heterogeneity of lesion appearances, and typically small publicly available datasets.\\

\noindent The autoPET/CT IV challenge directly addresses these challenges. In Task 2, the organizers provide a longitudinal whole-body CT dataset~\cite{kuestner2025longitudinalct}, consisting of 300 patients with baseline and follow-up images, accompanied by ground truth segmentation masks and localization information for each individual lesion. The challenge effectively poses the task of lesion tracking as a longitudinal promptable segmentation problem.\\

\noindent This formulation aligns with recent advances in medical image segmentation. First, after the introduction of the Segment Anything Model (SAM)~\cite{kirillov2023segment}, interactive methods also gained traction in the medical domain~\cite{wong2024scribbleprompt,wang2024sam,isensee2025nninteractive}. Second, longitudinal frameworks such as \textit{LongiSeg}~\cite{rokuss2024longitudinal} and \textit{LesionLocator}~\cite{rokuss2025lesionlocator} have demonstrated the benefits of incorporating temporal context for longitudinal image segmentation and lesion tracking.\\

\noindent In this work, we build upon these developments and propose a framework tailored to Task 2 of the autoPET/CT IV Challenge. Specifically, we extend the LongiSeg framework with promptable capabilities for point- and mask-based interactions. Furthermore, we employ large-scale pretraining on a synthetic longitudinal CT dataset to overcome limitations of the challenge dataset size. We systematically evaluate different design decisions, including model input, prompt representation and pretraining strategies, demonstrating the benefits of longitudinal pretraining for robust lesion tracking.

\section{Datasets}

\subsection{Training data}
The model is trained on the provided dataset~\cite{kuestner2025longitudinalct}. This dataset includes baseline and follow-up scans of 300 patients, including ground truth segmentations, lesion centers and propagated center locations in the follow-up for each individual lesion. Out of the 300 provided patients, 15 were excluded due to points being at the edge of scans, which likely represents cases, where the lesion location is not present on the follow-up.

\subsection{Pretraining data}
In addition to the provided dataset we utilized the \textit{LesionLocator} syntetic longitudinal CT dataset~\cite{rokuss2025lesionlocator} for pretraining of our model. This dataset includes real CT volumes of 2625 patients which are augmented using anatomy informed data augmentation~\cite{kovacs2023anatomy} to generate a synthetic baseline scan.

\section{Methods}

Our approach builds upon the recently proposed \textit{LongiSeg} framework~\cite{rokuss2024longitudinal}, which was originally developed for fully automatic longitudinal segmentation of aligned baseline–follow-up scans. In contrast, the more recent \textit{LesionLocator}~\cite{rokuss2025lesionlocator} is a promptable framework that enables lesion tracking starting from a single baseline prompt by propagating it to future timepoints. While LesionLocator is well suited for the general setting of lesion tracking without follow-up annotations, the autoPET/CT IV challenge provides prompts in both baseline and follow-up scans. This renders prompt propagation unnecessary and makes the LongiSeg formulation, with direct use of predefined patches around the provided prompts, the more appropriate choice for this task.

\subsection{Network architecture}

\paragraph{Backbone and longitudinal inputs:} We utilize the powerful ResEncL~\cite{isensee2024nnu} preset as a model backbone, which generates a deep U-Net architecture with multiple residual blocks at each encoder layer. Longitudinal inputs are aligned via the given center locations and concatenated along the channel dimension as input to the model.

\paragraph{Prompt representation:} We follow the same strategy we use for longitudinal inputs of providing both point as well as mask prompts via additional input channels to the U-Net. Point prompts are represented as Gaussian blobs, which we rescale to unit intensity at the center to approximately match the intensity normalization of other input channels. Initial experiments showed significant gains with respect to normalizing the Gaussians to unit volumes.

\subsection{Data sampling:} During training single lesions are sampled from random patients. Baseline and follow-up scans are aligned via the given center locations and both scans are randomly shifted by up to 4 voxels in each direction. Patches from both scans are extracted, placing the center at a random location in the inner half of each patch. During inference the center locations are perfectly aligned and patches are extracted such that the center lies directly in the middle of the patches to guarantee best possible performance.

\subsection{Inference}

During inference the network does a single forward pass per lesion and either exports each prediction separately or merges the predictions into a single multilabel segmentation map.

\section{Results}

\begin{table}[htbp]
\begin{center}
    \caption{Results from the five-fold cross-validation. Predictions are merged according to the ground truths and metrics are calculated for each merged lesion pair and averaged over patients. Fold 4 is excluded from the mean as it was unstable during training, full results can be found in the appendix~\ref{tab:results_full}.
    }
    \label{tab:results}
        \setlength{\tabcolsep}{4pt} 
        \begin{tabular}{l|ccc}
        Setting & Dice$\uparrow$ & FNvol$\downarrow$ & FPvol$\downarrow$ \\ 
        \hline
        Cross Sectional + Point & 53.06 & 1532 & \textbf{113} \\
        Cross Sectional + Mask & 56.64 & 808 & 195 \\
        Longitudinal + Mask + Point & 55.78 & 1005 & 355 \\
        \hline
        Longitudinal Batch Size 2 & 58.08 & 736 & 374 \\
        \hline
        Pretrained nnInteractive Weights & 59.28 & 763 & 630 \\
        Pretrained LesionLocator Weights & 59.57 & 689 & 626 \\
        Pretrained Cross Sectional + Mask & 62.27 & \textbf{266} & 372 \\
        Pretrained Synth. Longitudinal Data  & 62.89 & 366 & 177 \\
        \hline
        Final & \textbf{63.71} & 343 & 144 \\
        \end{tabular}
    \end{center}
\end{table}

\begin{figure}[htbp]
    \centering
    \begin{subfigure}[t]{0.45\textwidth}
        \centering
        \includegraphics[width=\textwidth]{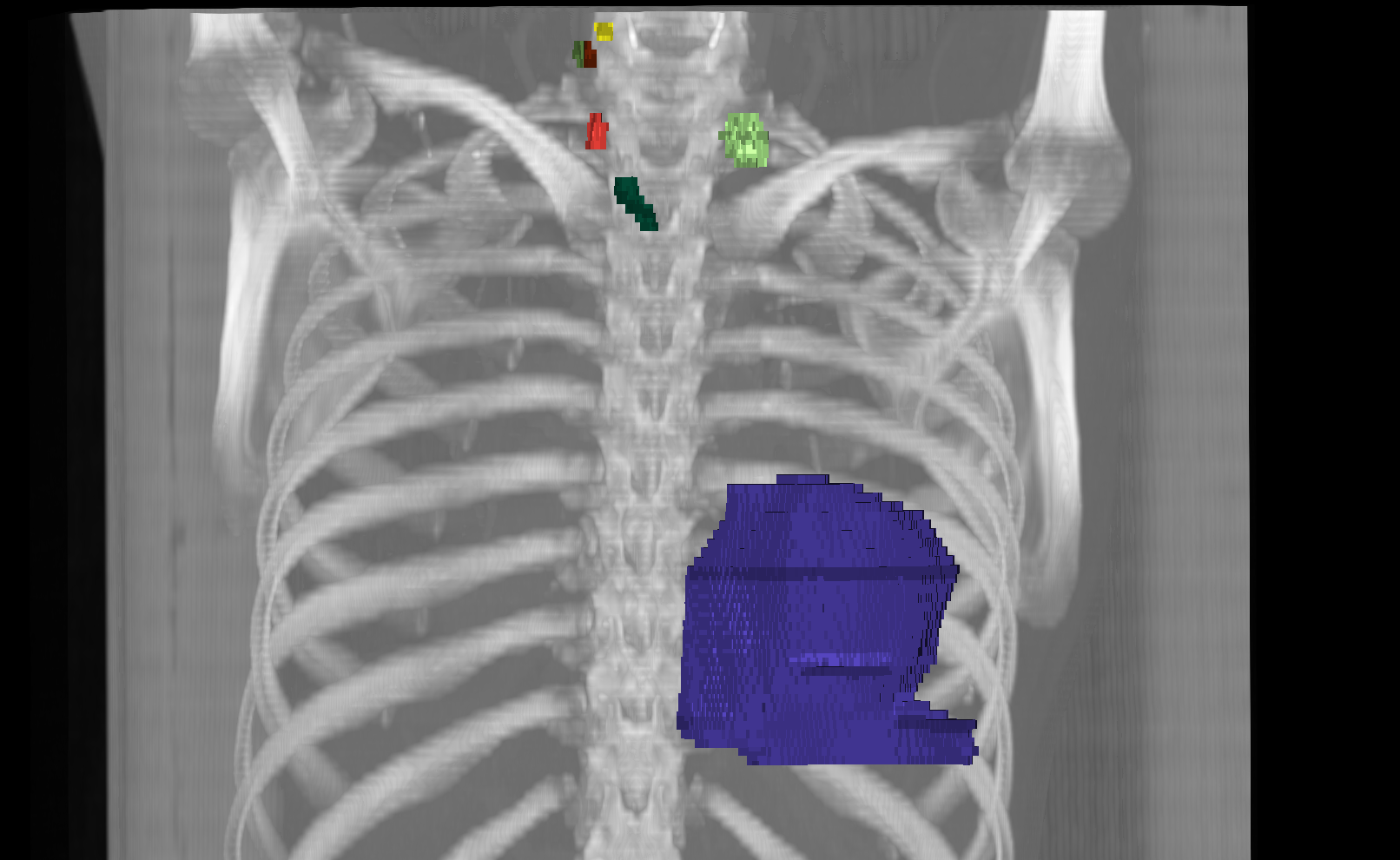}
        \caption{Case 1 – Ground Truth}
    \end{subfigure}
    \begin{subfigure}[t]{0.45\textwidth}
        \centering
        \includegraphics[width=\textwidth]{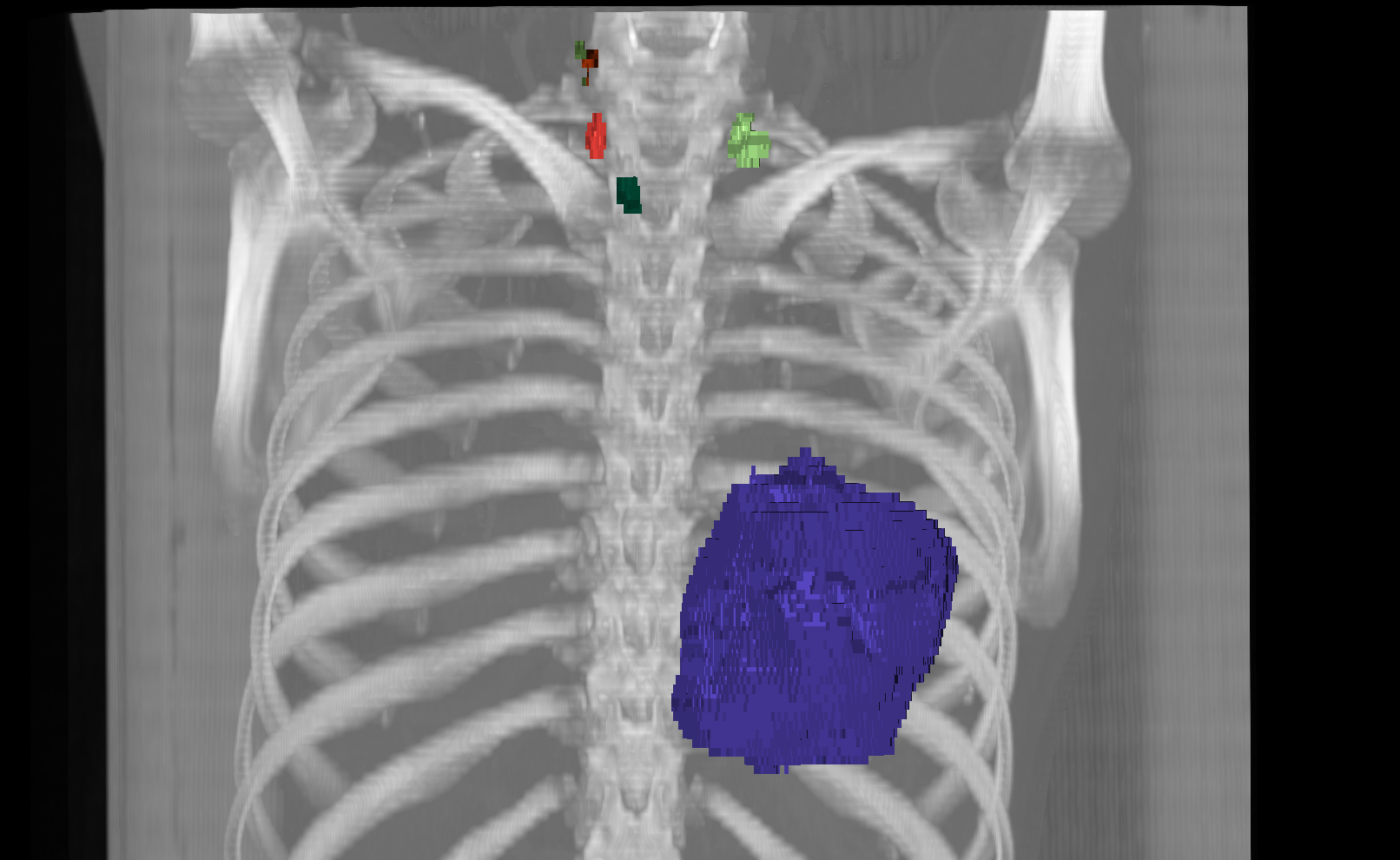}
        \caption{Case 1 - Prediction}
    \end{subfigure}
    \begin{subfigure}[t]{0.45\textwidth}
        \centering
        \includegraphics[width=\textwidth]{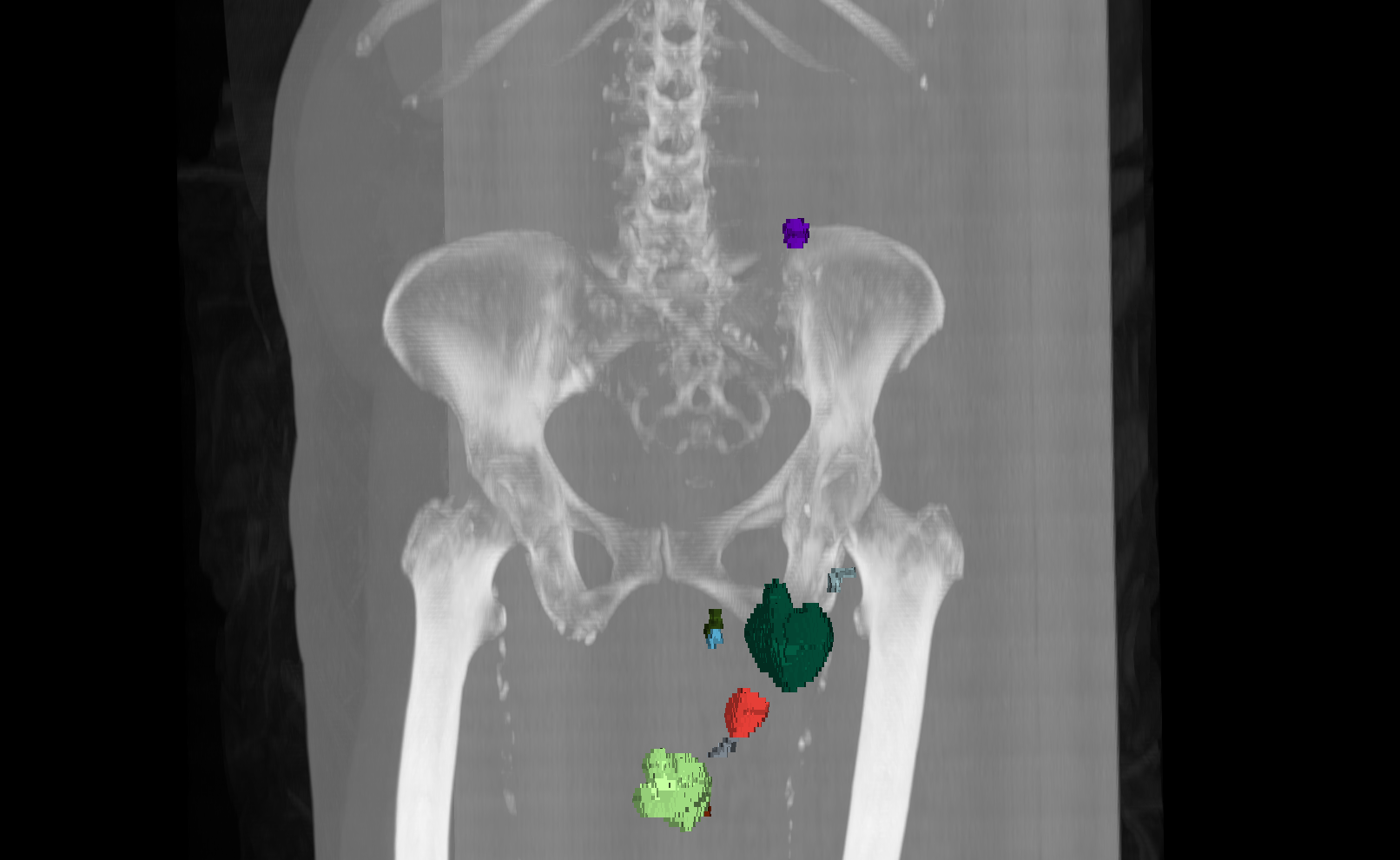}
        \caption{Case 2 - Ground Truth}
    \end{subfigure}
    \begin{subfigure}[t]{0.45\textwidth}
        \centering
        \includegraphics[width=\textwidth]{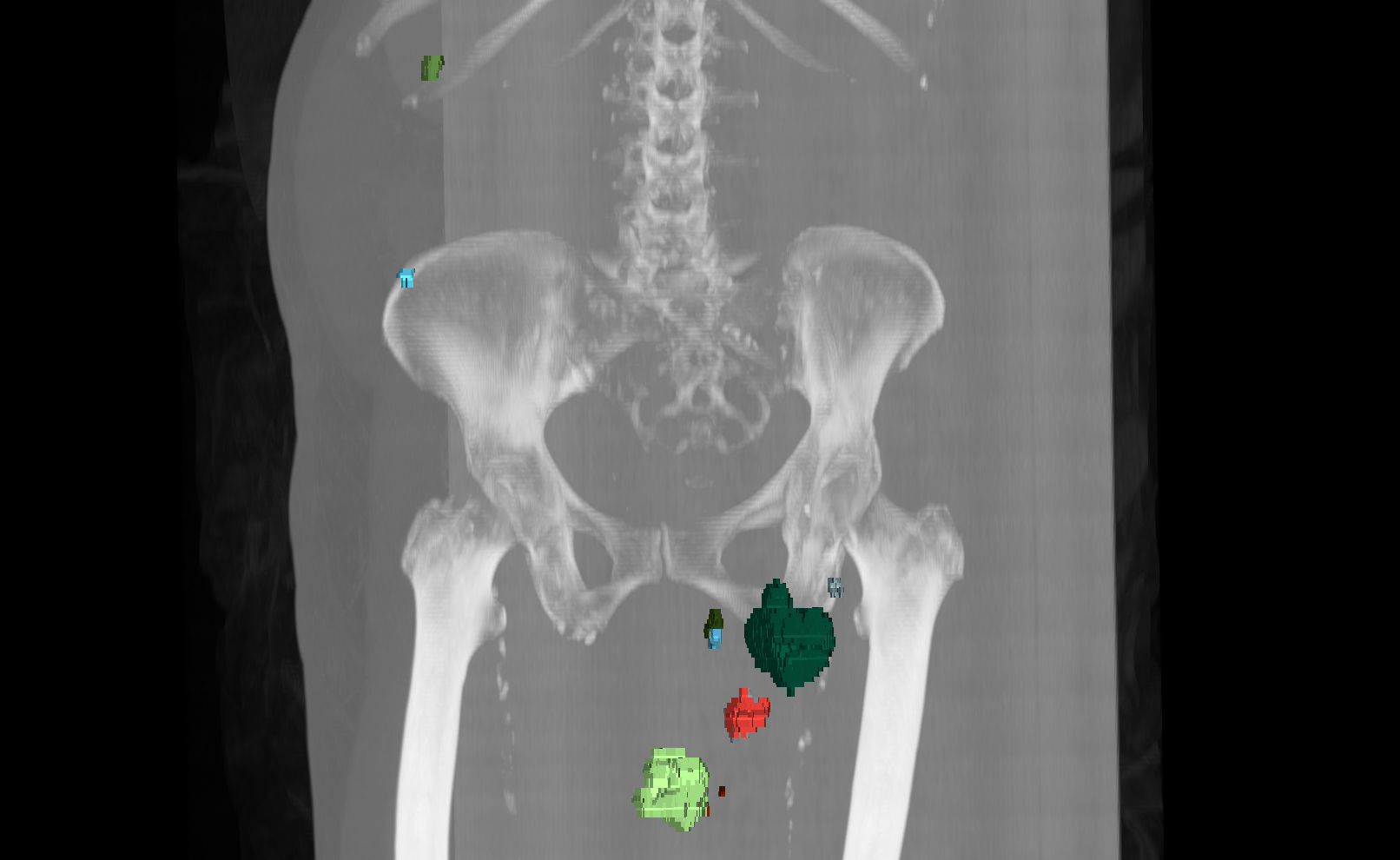}
        \caption{Case 2 - Prediction}
    \end{subfigure}
    \caption{Qualitative results of our final model on two cases from the cross-validation. The top row shows Case 1 (ID: \texttt{013d407166}), and the bottom row shows Case 2 (ID: \texttt{31fefc0e57}). Each pair displays the ground truth (left) and model prediction (right). The model tracks and segments most lesions accurately, with inaccuracies mainly in boundary regions.}
    \label{fig:qual_results}
\end{figure}

Table~\ref{tab:results} reports the results of our five-fold cross-validation experiments, evaluating Dice, false negative volume (FNvol) and false positive volume (FPvol). Different ablations are shown, including different inputs, batch sizes  and pretrainings. Without pretraining, the model trained only on the current image and the prior mask performs best, indicating that it is not able to learn the full longitudinal context from the provided dataset alone. However, pretraining the models on the large synthetic dataset enables the model to properly utilize the information from the previous timepoint, outperforming the single timepoint solution by 0.6 dice points. Experiments with different batch sizes show that a smaller batch size of 2 performs noticeably better than our default batch size of 4. While an initialization with the pretrained LesionLocator~\cite{rokuss2025lesionlocator} checkpoint performs slightly better than using nnInteractive~\cite{isensee2025nninteractive} as an initialization, both do not come close to longitudinal pretraining on the synthetic dataset. Scaling up this pretraining with a larger batch size gives a further boost in performance, improving the dice score, while simultaneously reducing both false negative and false positive volumes. Figure~\ref{fig:qual_results} shows qualitative results of our best model on two cases from the cross-validation, underlining its strong performance.

\paragraph{Test Set Submission:} For the final submission we ensemble the five folds from the best performing model on the cross-validation. This is pretrained with a large batch size on the synthetic longitudinal dataset and gets both timepoints, the prior segmentation and the point prompt as inputs via channel concatenation.

\section{Conclusion}

This paper presents our contribution to Task 2 of the autoPET/CT IV. We extended the LongiSeg framework to incorporate point- and mask-based prompts to enable tracking lesions across different timepoints. Additionally, we use a large synthetic longitudinal dataset to pretrain our model, which significantly improves performance upon training from scratch. Our final submission is based on a five-fold ensemble with large-scale pretraining on the synthetic dataset. Our results demonstrate the effectiveness of incorporating longitudinal information together with point- and mask-based prompts within the LongiSeg framework for efficient longitudinal lesion tracking.

\subsubsection*{Acknowledgements}
The present contribution is supported by the Helmholtz Association under the joint research school ”HIDSS4Health – Helmholtz Information and Data Science School for Health”. This work was partly funded by Helmholtz Imaging (HI), a platform of the Helmholtz Incubator on Information and Data Science.

\bibliographystyle{splncs04}
\bibliography{bib}

\newpage

\appendix

\section{Full model results}

\begin{table}[htbp]
\begin{center}
    \caption{Results from the five-fold cross-validation. Predictions are merged according to the ground truths and metrics are calculated for each merged lesion pair and averaged over patients.\\
    *: Training on Fold 4 collapsed with resulting dice of 0
    }
    \label{tab:results_full}
        \setlength{\tabcolsep}{4pt} 
        \begin{tabular}{l|ccc}
        Setting & Dice$\uparrow$ & FNvol$\downarrow$ & FPvol$\downarrow$ \\ 
        \hline
        Cross Sectional + Point & 52.81 & 1353 & \textbf{107} \\
        Cross Sectional + Mask & 56.28 & 771 & 266 \\
        Longitudinal + Mask + Point & 55.62 & 902 & 394 \\
        \hline
        Longitudinal Batch Size 2* & 46.47 & 2456 & 299 \\
        \hline
        Pretrained nnInteractive & 58.90 & 735 & 616 \\
        Pretrained LesionLocator & 58.96 & 678 & 608 \\
        Pretrained Cross Sectional + Mask & 61.20 & \textbf{345} & 346 \\
        Pretrained Longitudinal & 61.94 & 404 & 243 \\
        \hline
        Final & \textbf{62.49} & 386 & 132 \\
        \end{tabular}
    \end{center}
\end{table}

\end{document}